\documentclass{article}

\begin{document}

\thispagestyle{empty}

\begin{center}
{\Large 
Topological Charge of Noncommutative ADHM Instanton}
\end{center}

\vspace*{2cm} 

\begin{center}
\noindent {\large Yu Tian} \vspace{5mm} \noindent \hspace{0.7cm}
\parbox{120mm}{\it
School of Physics, Peking University, Beijing 100871, China
\\
E-mail: {\tt phytian@yahoo.com}
 }
\end{center}

\vspace{5mm} 

\begin{center}
\noindent {\large Chuan-Jie Zhu} \vspace{5mm} \noindent \hspace{0.7cm}
\parbox{120mm}{\it
Institute of Mathematics, Henan University, Kaifeng 475001
\\
and
\\
Institute for Theoretical Physics, Chinese Academy of Sciences
\\
P. O. Box 2735, Beijing 100080
\\
E-mail: {\tt zhucj@itp.ac.cn}
 }
\end{center}

\vspace{5mm} 

\begin{center}
\noindent {\large Xing-Chang Song} \vspace{5mm} \noindent
\hspace{0.7cm}
\parbox{120mm}{\it
Institute for Theoretical Physics, Chinese Academy of Sciences
\\
and
\\
School of Physics, Peking University, Beijing 100871, China
\\
E-mail: {\tt songxc@pku.edu.cn}
 }
\end{center}

\vspace{2cm}

\begin{center}
\textbf{Abstract}
\end{center}
We analytically calculate the topological charge of the noncommutative ADHM $%
U(N)$ $k$-instanton using the Corrigan's identity and find that the result
is exactly the instanton number $k$.

\newpage

\section{Introduction}

Instanton solutions in gauge field theory are interesting in both physics and mathematics. In ordinary (commutative) case, it is well-known that the topological charge of a $U(N)$ $%
k$-instanton is equal to $k$. But in noncommutative gauge field
theory the relation between the topological charge and the
instanton number is not clear. Using the Corrigan's identity
\cite{Corrigan}, a well-known argument was presented \cite{Paperd}
that the topological charge of noncommutative $U(N)$ $k$-instanton
is also equal to $k$. And in all the known cases, numerical calculations \cite%
{Paperd}\cite{TianZhu} or some analytic method \cite{Calculus} shows that
this conclusion is correct. But there is no strict proof for general $U(N)$ $%
k$-instanton cases yet.

In this paper we will analytically calculate the topological
charge of noncommutative ADHM $U(N)$ $k$-instanton in the case of
non-degenerate $\theta$. We also make use of the Corrigan's
identity, but we work in the Fock space representation. This is
the point why we can avoid the possible singularity when using the
usual (function star product) representation. To achieve this,
however, a little mathematics about the Hilbert space will be
needed. We note that a recent paper \cite{charge} also considered
this problem and got the same result. But their method seems much
more complicated than ours, and they do not include the
$\theta_2<0$ (see below for our notations) case.

As we have known, the limit of $\theta_2\rightarrow 0$ is
singular. So the degenerate case, $\theta_2=0$, must be considered
separately. But our method fails in this case because we can not
introduce a Fock space representation for $z_2$ and $\bar{z}_2$.
This problem is left for future works.

The organization of this paper is as follows. In section 2 we
recall briefly the noncommutative $\bf{R}^4$ and noncommutative
instanton. In section 3 we review the famous ADHM construction,
first in the commutative case, then in noncommutative case. In
section 4 we calculate the topological charge of noncommutative
instantons. Finally we present in the appendix the mathematical
foundation concerning the non-singularity of our method.

\section{$\mathbf{{R}_{\mathrm{NC}}^{4}}$ and the (anti-)self-dual equations}

First let us recall briefly the noncommutative $\mathbf{{R}^{4}}$ and set
our notations\footnote{%
For general reviews on noncommutative geometry and field theory, see, for
example, \cite{Paperc, Reviewa, Reviewb, Reviewc}.}. For a general
noncommutative $\mathbf{{R}^{4}}$ we mean a space with (operator)
coordinates $x^{m}$, $m=1,\cdots ,4$, which satisfy the following relations:
\begin{equation}
\lbrack x^{m},x^{n}]=i\theta ^{mn},
\end{equation}%
where $\theta ^{mn}$ are real constants. If we assume the standard
(Euclidean) metric for the noncommutative $\mathbf{{R}^{4}}$, we can use the
orthogonal transformation with positive determinant to change $\theta ^{mn}$
into the following standard form:
\begin{equation}
(\theta ^{mn})=\left(
\begin{array}{cccc}
0 & \theta ^{12} & 0 & 0 \\
-\theta ^{12} & 0 & 0 & 0 \\
0 & 0 & 0 & \theta ^{34} \\
0 & 0 & -\theta ^{34} & 0%
\end{array}%
\right) ,  \label{theta}
\end{equation}%
where $\theta ^{12}>0$ and $\theta ^{12}+\theta ^{34}\geq 0$. By using this
form of $\theta ^{mn}$, the only non-vanishing commutators are as follows:
\begin{equation}
\lbrack x^{1},x^{2}]=i\theta ^{12},\qquad \lbrack x^{3},x^{4}]=i\theta ^{34},
\end{equation}%
and other twos obtained by using the anti-symmetric property of the
commutators. Introducing complex coordinates:
\begin{equation}
\begin{array}{rl}
z_{1}=x^{2}+ix^{1}, & \bar{z}_{1}=x^{2}-ix^{1}, \\
z_{2}=x^{4}+ix^{3}, & \bar{z}_{2}=x^{4}-ix^{3},%
\end{array}
\label{complex}
\end{equation}%
the non-vanishing commutation relations are
\begin{equation}
\lbrack \bar{z}_{1},z_{1}]=2\theta ^{12}\equiv \theta _{1},\quad \lbrack
\bar{z}_{2},z_{2}]=2\theta ^{34}\equiv \theta _{2}.  \label{z commutator}
\end{equation}

By a noncommutative gauge field $A_m$ we mean an operator valued field. The
(anti-hermitian) field strength $F_{mn}$ is defined similarly as in the
commutative case:
\begin{equation}
F_{mn}=\hat\partial_{[m} A_{n]} + A_{[m} A_{n]} \equiv \hat\partial_m A_n -
\hat\partial_n A_m + [A_m, A_n],  \label{F by A}
\end{equation}
where the derivative operator $\hat\partial_m$ is defined as follows:
\begin{equation}
\hat\partial_m f \equiv - i \theta_{mn} [x^n, f],
\end{equation}
where $\theta_{mn}$ is the inverse of $\theta^{mn}$. For our standard form (%
\ref{theta}) of $\theta^{mn}$ we have
\begin{equation}
\hat\partial_1 f = {\frac{i }{\theta^{12}}} [x^2,f], \qquad \hat\partial_2 f
= - {\frac{i }{\theta^{12}}} [x^1,f],
\end{equation}
which can be expressed by the complex coordinates (\ref{complex}) as
follows:
\begin{equation}
\partial_1 f \equiv \hat\partial_{z_1}f = {\frac{1}{\theta_1}}[\bar z_1, f],
\qquad \bar\partial_1 f \equiv \hat\partial_{\bar z_1}f = - {\frac{1}{%
\theta_1}}[z_1, f],
\end{equation}
and similar relations for $x^{3,4}$ and $z_2,\bar{z}_2$.

For a general metric $g_{mn}$ the instanton equations are
\begin{equation}  \label{instanton}
F_{mn}=\pm\frac{\epsilon^{pqrs}}{2\sqrt{g}}g_{mp}g_{nq}F_{rs},
\end{equation}
and the solutions are known as self-dual (SD, for ``+'' sign) and
anti-self-dual (ASD, for ``$-$'' sign) instantons. Here $\epsilon^{pqrs}$ is
the totally anti-symmetric tensor ($\epsilon^{1234}=1$ etc.) and $g$ is the
metric. We will take the standard metric $g_{mn}=\delta_{mn}$ and take the
noncommutative parameters $\theta_{1,2}$ as free parameters. We also note
that the notions of self-dual and anti-self-dual are interchanged by a
parity transformation. A parity transformation also changes the sign of $%
\theta^{mn}$. In the following discussion we will consider only the ASD
instantons. So we should not restrict $\theta_2$ to be positive.

\section{Instantons in Noncommutative Gauge Theory}

\subsection{ADHM construction for ordinary gauge theory}

For ordinary gauge theory all the (ASD) instanton solutions are obtained by
ADHM (Atiyah-Drinfeld-Hitchin-Manin) construction \cite{ADHM}. In this
construction we introduce the following ingredients (for $U(N)$ gauge theory
with instanton number $k$):

\begin{itemize}
\item complex vector spaces $V$ and $W$ of dimensions $k$ and $N$,

\item $k\times k$ matrix $B_{1,2}$, $k\times N$ matrix $I$ and $N\times k$
matrix $J$,

\item the following quantities:
\begin{eqnarray}
\mu _{r} &=&[B_{1},B_{1}^{\dagger }]+[B_{2},B_{2}^{\dagger }]+I\,I^{\dagger
}-J^{\dagger }J,  \label{ADHM1} \\
\mu _{c} &=&[B_{1},B_{2}]+I\,J.  \label{ADHM2}
\end{eqnarray}
\end{itemize}

The claim of ADHM is as follows:

\begin{itemize}
\item Given $B_{1,2}$, $I$ and $J$ such that $\mu_r=\mu_c=0$, an ASD gauge
field can be constructed;

\item All ASD gauge fields can be obtained in this way.
\end{itemize}

It is convenient to introduce a quaternionic notation for the 4-dimensional
Euclidean space-time indices:
\begin{equation}
x\equiv x^{n}\sigma _{n},\qquad \bar{x}\equiv x^{n}\bar{\sigma}_{n},
\end{equation}%
where $\sigma _{n}=(i\vec{\tau},1)$ and $\tau ^{c}$, $c=1,2,3$ are the three
Pauli matrices, and the conjugate matrices $\bar{\sigma}_{n}=\sigma
_{n}^{\dag }=(-i\vec{\tau},1)$. In terms of the complex coordinates (\ref%
{complex}) we have
\begin{equation}
(x_{\alpha \dot{\alpha}})=\left(
\begin{array}{cc}
z_{2} & z_{1} \\
-\bar{z}_{1} & \bar{z}_{2}%
\end{array}%
\right) ,\qquad (\bar{x}^{\dot{\alpha}\alpha })=\left(
\begin{array}{cc}
\bar{z}_{2} & -z_{1} \\
\bar{z}_{1} & z_{2}%
\end{array}%
\right) .  \label{quatern}
\end{equation}%
Then the basic object in the ADHM construction is the $(N+2k)\times 2k$
matrix $\Delta $ which is linear in the space-time coordinates:
\begin{equation}
\Delta =a+b\bar{x},  \label{Delta}
\end{equation}%
where the constant matrices
\begin{equation}
a=\left(
\begin{array}{cc}
I^{\dag } & J \\
B_{2}^{\dagger } & -B_{1} \\
B_{1}^{\dagger } & B_{2}%
\end{array}%
\right) ,\quad b=\left(
\begin{array}{cc}
0 & 0 \\
1 & 0 \\
0 & 1%
\end{array}%
\right) .
\end{equation}

Consider the conjugate operator of $\Delta$:
\begin{equation}
\Delta^\dagger = a^\dag +x b^\dag = \left( \matrix{I & B_2 + z_2 & B_1 + z_1
\cr J^\dagger & -B_1^\dagger -\bar z_1 & B_2^\dagger + \bar z_2} \right).
\end{equation}
It is easy to check that the ADHM equations (\ref{ADHM1}) and (\ref{ADHM2})
are equivalent to the so-called factorization condition:
\begin{equation}  \label{factorize}
\Delta^\dagger\Delta=\left(\matrix{f^{-1} & 0 \cr 0 & f^{-1}}\right),
\end{equation}
where $f(x)$ is a $k\times k$ hermitian matrix. From the above
condition we can construct a hermitian projection operator $P$ as
follows:\footnote{We use the following abbreviation for
expressions with $f$: $$\Delta f \Delta^\dag \equiv \Delta
\left(\matrix{f & 0 \cr 0 & f}\right) \Delta^\dag =\Delta
(f\otimes 1_2)\Delta^\dag.$$}
\begin{eqnarray}
P&=&\Delta f\Delta^\dag, \cr P^2&=&\Delta f f^{-1}f\Delta^\dag=P.
\end{eqnarray}

Obviously, the null-space of $\Delta^\dagger(x)$ is of $N$ dimension for
generic $x$. The basis vector for this null-space can be assembled into an $%
(N+2k)\times N$ matrix $U(x)$:
\begin{equation}
\Delta^\dag U=0,
\end{equation}
which can be chosen to satisfy the following orth-normalization condition:
\begin{equation}  \label{normal}
U^\dag U=1.
\end{equation}
The above orth-normalization condition guarantees that $UU^\dag$
is also a hermitian projection operator. Now it can be proved (see
\cite{TianZhu2}) that the completeness relation\footnote{This
relation in noncommutative case was first showed in
\cite{Paperb}.}
\begin{equation}  \label{complete}
P+UU^\dag=1
\end{equation}
holds if $U$ contains the whole null-space of $\Delta^\dagger$. In
other words, this completeness relation requires that $U$ consists
of all the zero modes of $\Delta^\dagger$.

The (anti-hermitian) gauge potential is constructed from $U$ by the
following formula:
\begin{equation}
A_m= U^\dag\partial_m U.
\end{equation}
Substituting this expression into (\ref{F by A}), we get the following field
strength:
\begin{eqnarray}
F_{mn}&=&\partial_{[m}(U^\dag\partial_{n]}U)
+(U^\dag\partial_{[m}U)(U^\dag\partial_{n]}U)
=\partial_{[m}U^\dag(1-UU^\dag)\partial_{n]}U  \nonumber \\
&=&\partial_{[m}U^\dag\Delta f\Delta^\dag\partial_{n]}U
=U^\dag\partial_{[m}\Delta f\partial_{n]}\Delta^\dag U =U^\dag
b\bar\sigma_{[m}\sigma_{n]}f b^\dag U  \nonumber \\
&=& 2i\bar\eta^c_{mn}U^\dag b(\tau^c f)b^\dag U.  \label{F by U}
\end{eqnarray}
Here $\bar\eta^a_{ij}$ is the standard 't Hooft $\eta$-symbol, which is
anti-self-dual:
\begin{equation}
\frac{1}{2}\epsilon_{ijkl}\bar\eta^a_{kl}=-\bar\eta^a_{ij}.
\end{equation}

\subsection{Noncommutative ADHM construction}

The above construction has been extended to noncommutative gauge theory \cite%
{Schwarz}. We recall this construction briefly here. By introducing the same
data as above but considering the $z_i$'s as noncommutative we see that the
factorization condition (\ref{factorize}) still gives $\mu_c=0$, but $\mu_r$
no longer vanishes. It is easy to check that the following relation holds:
\begin{equation}
\mu_r=\zeta\equiv\theta_1+\theta_2.
\end{equation}
In this case the two ADHM equations (\ref{ADHM1}) and (\ref{ADHM2}) can be
combined into one \cite{Paperd}:
\begin{equation}  \label{ADHM}
\tau^{c\dot\alpha}{}_{\dot\beta}(\bar a^{\dot\beta}
a_{\dot\alpha})_{ij}=\delta_{ij}\delta^{c3}\zeta.
\end{equation}

As studied mathematically by various people (see, for example, the lectures
by H. Nakajima \cite{Nakajima}), the moduli space of the noncommutative
instantons is better behaved than their commutative counterpart. In the
noncommutative case the operator $\Delta^\dagger\Delta$ always has maximum
rank (see \cite{Reviewa}).

Though there is no much difference between the noncommutative ADHM
construction and the commutative one, we should study the noncommutative
case in more detail. In order to study the instanton solution precisely, we
use a Fock space representation as follows ($n_{1},n_{2}\geq 0$):
\begin{eqnarray}
z_{1}|n_{1},n_{2}\rangle &=&\sqrt{\theta _{1}}\sqrt{n_{1}+1}%
|n_{1}+1,n_{2}\rangle , \\
\quad \bar{z}_{1}|n_{1},n_{2}\rangle &=&\sqrt{\theta _{1}}\sqrt{n_{1}}%
|n_{1}-1,n_{2}\rangle ,
\end{eqnarray}%
by using the commutation relation (\ref{z commutator}). Similar expressions
for $z_{2}$ and $\bar{z}_{2}$ also apply (but paying a little attention to
the sign of $\theta _{2}$ which is not restricted to be positive). In this
representation the $z_{i}$'s are infinite-dimensional matrices, and so are
the operator $\Delta $, $\Delta ^{\dag }$ etc. Because of infinite
dimensions are involved we can not determine the dimension of null-space of $%
\Delta ^{\dag }$ straightforwardly from the difference of the numbers of its
rows and columns. But it turns out that $\Delta ^{\dag }$ also has infinite
number of zero modes, and they can be arranged into an $(N+2k)\times N$
matrix with entries from the (noncommutative) algebra generated by the
coordinates, which resembles the commutative case.

In the Fock space representation we have the operator trace%
\begin{equation}
\mathrm{Tr}_\mathcal{F}\mathcal{O}\equiv
\sum_{n_{1},n_{2}=0}^{\infty }\langle
n_{1},n_{2}|\mathcal{O}|n_{1},n_{2}\rangle . \label{trace}
\end{equation}%
The integral on $\mathbf{{R}_{\mathrm{NC}}^{4}}$ is defined as%
\begin{equation}
\int d^{4}x=(2\pi )^{2}\sqrt{\det \theta
}\mathrm{Tr}_\mathcal{F}=\pi ^{2}|\theta _{1}\theta
_{2}|\mathrm{Tr}_\mathcal{F}.
\end{equation}

\section{Calculation of the Topological Charge}

The topological charge of a $U(N)$ ASD gauge field $F$ is given by%
\begin{equation}
Q=-\frac{1}{8\pi ^{2}}\int \mathrm{Tr}_{N}(F\wedge F)=\frac{1}{16\pi ^{2}}%
\int d^{4}x\mathrm{Tr}_{N}(F_{mn}F_{mn}),
\end{equation}%
where in the second equation we have used the anti-self-duality of $F$. And
the ASD field strength is given by (\ref{F by U}):%
\begin{equation}
F_{mn}=U^{\dag }b\bar{\sigma}_{[m}\sigma _{n]}fb^{\dag }U.
\end{equation}%
Then the Corrigan's identity says%
\begin{equation}
\mathrm{Tr}_{N}(F_{mn}F_{mn})=\frac{1}{2}\partial _{n}\partial _{n}\mathrm{Tr%
}[\sigma _{m}b^{\dag }(2-\Delta f\Delta ^{\dag })b\bar{\sigma}_{m}f].
\end{equation}%
Use the fact that%
\begin{equation}
\partial _{n}\partial _{n}=4\partial _{\alpha }\bar{\partial}_{\alpha
},\quad \alpha =1,2,
\end{equation}%
we can have%
\begin{equation}
\mathrm{Tr}_{N}(F_{mn}F_{mn})=8\partial _{\alpha }\bar{\partial}_{\alpha }%
\mathrm{Tr}[b^{\dag }(2-\Delta f\Delta ^{\dag })bf].
\end{equation}%
So the topological charge is expressed as follows:%
\begin{eqnarray}
Q &=&\frac{1}{2\pi ^{2}}\int d^{4}x\partial _{\alpha }\bar{\partial}_{\alpha
}\mathrm{Tr}[b^{\dag }(2-\Delta f\Delta ^{\dag })bf]  \nonumber \\
&=&\frac{|\theta _{1}\theta _{2}|}{2}\mathrm{Tr}_\mathcal{F}\partial _{\alpha }\bar{%
\partial}_{\alpha }\mathrm{Tr}[(2-b^{\dag }\Delta f\Delta ^{\dag }b)f].
\end{eqnarray}

First consider the $\theta _{2}>0$ case. We note for an operator
$\mathcal{O}_{1}$ (or $\mathcal{O}_{2}$)
\begin{eqnarray}
\langle n_{1},n_{2}|\partial _{1}\mathcal{O}_{1}|n_{1},n_{2}\rangle
&=&\langle n_{1},n_{2}|\frac{\bar{z}_{1}\mathcal{O}_{1}-\mathcal{O}_{1}\bar{z%
}_{1}}{\theta _{1}}|n_{1},n_{2}\rangle  \nonumber \\
&=&\theta _{1}^{-1/2}(\sqrt{n_{1}+1}\langle n_{1}+1,n_{2}|\mathcal{O}%
_{1}|n_{1},n_{2}\rangle \label{diff2}\\
&&-\sqrt{n_{1}}\langle n_{1},n_{2}|\mathcal{O}_{1}|n_{1}-1,n_{2}\rangle)
\nonumber
\end{eqnarray}%
and similar expression for $\langle n_{1},n_{2}|\partial _{2}\mathcal{O}%
_{2}|n_{1},n_{2}\rangle $, provided the terms on the right hand
side are well-defined. Because the above expressions have the form
of difference, we can conclude that if we sum $\langle
n_{1},n_{2}|\partial _{\alpha }\mathcal{O}_{\alpha
}|n_{1},n_{2}\rangle $ on a finite region of the $n_{1}n_{2}$
plane the terms corresponding to the interior will be cancelled
out and we will get the flux of $\mathcal{O}_{\alpha }$ across the
boundary when considering $\mathcal{O}_{\alpha }$ as a
2-dimensional vector field, which resembles the Stoke's theorem.
Specially, we have for the
rectangle region%
\begin{eqnarray}
\sum_{n_{1},n_{2}=0}^{N_{1},N_{2}}\langle n_{1},n_{2}|\partial _{\alpha }%
\mathcal{O}_{\alpha }|n_{1},n_{2}\rangle &=&\sqrt{\frac{N_{1}+1}{\theta _{1}}%
}\sum_{n_{2}=0}^{N_{2}}\langle N_{1}+1,n_{2}|\mathcal{O}_{1}|N_{1},n_{2}%
\rangle  \label{Stokes} \\
&&+\sqrt{\frac{N_{2}+1}{\theta _{2}}}\sum_{n_{1}=0}^{N_{1}}\langle
n_{1},N_{2}+1|\mathcal{O}_{2}|n_{1},N_{2}\rangle ,  \nonumber
\end{eqnarray}%
for the contributions of $n_{1}=0$ and $n_{2}=0$ boundaries vanish.

Turn to the calculation of the topological charge, it is easy to see that in
order to apply (\ref{Stokes}) we can first introduce a truncation of $%
n_{1}=N_{1}$ and $n_{2}=N_{2}$ when doing the trace (\ref{trace}) and then
take the limit of $N_{1},N_{2}\rightarrow \infty $. In this case%
\begin{equation}
\mathcal{O}_{\alpha }=\bar{\partial}_{\alpha }\mathcal{O},\
\mathcal{O\equiv }(2-b^{\dag }\Delta f\Delta ^{\dag }b)f, \label{O
by f}
\end{equation}%
so (\ref{Stokes}) can be deduced further, for%
\begin{eqnarray}
\langle
n_{1}+1,n_{2}|\bar{\partial}_{1}\mathcal{O}|n_{1},n_{2}\rangle
&=&\langle n_{1}+1,n_{2}|\frac{\mathcal{O}z_{1}-z_{1}\mathcal{O}}{\theta _{1}%
}|n_{1},n_{2}\rangle  \nonumber \\
&=&\sqrt{\frac{n_{1}+1}{\theta _{1}}}(\langle n_{1}+1,n_{2}|\mathcal{O}%
|n_{1}+1,n_{2}\rangle  \label{diff} \\
&&-\,\langle n_{1},n_{2}|\mathcal{O}|n_{1},n_{2}\rangle).
\nonumber
\end{eqnarray}%
and similar expression for $\langle n_{1},n_{2}+1|\bar{\partial}_{2}\mathcal{%
O}|n_{1},n_{2}\rangle $. These equations imply that if $\langle
n_{1},n_{2}|\mathcal{O}|n_{1},n_{2}\rangle$ is well-defined for
any $n_{1},n_{2}$, the right hand side of (\ref{diff2}) is
well-defined. We have proved in appendix \ref{append} that the
operator $\mathcal{O}$ in (\ref{O by f}) is well-defined on any
states of the Hilbert space, which is much stronger than the
non-singularity of the diagonal elements $\langle
n_{1},n_{2}|\mathcal{O}|n_{1},n_{2}\rangle$.

In order to study the asymptotic behavior of $\mathcal{O}$ we recall%
\begin{equation}
f =(\Delta ^{\dagger }\Delta )^{-1}=(z_{\alpha }\bar{z}_{\alpha
}+B_\alpha^\dag z_{\alpha }+B_\alpha\bar{z}_{\alpha }+B_\alpha
B_\alpha^\dag+II^\dag)^{-1},
\end{equation}%
so for large $n_{1}$ or $n_{2}$\footnote{%
Here and in the following when $|n_{1},n_{2}\rangle $ acted on by matrix
operators we always mean $|n_{1},n_{2}\rangle \otimes 1_{2k}$.}%
\begin{eqnarray}
f|n_{1},n_{2}\rangle &=&(\theta_\beta n_\beta)^{-1}|n_{1},n_{2}\rangle +(\theta_\beta n_\beta)^{-2}c|n_{1},n_{2}\rangle \nonumber\\
&&+\,(\theta_\beta
n_\beta)^{-2}(b^+_\alpha\sqrt{n_\alpha}|n+\rangle_\alpha+b^-_\alpha\sqrt{n_\alpha}|n-\rangle_\alpha) \nonumber\\
&&+\,O(n^{-3})|n_{1},n_{2}\rangle+O(n^{-5/2})|n\pm\rangle \nonumber\\
&&+\,O(n^{-2})|\overline{n_{1},n_{2}}\rangle,\quad
n\equiv\max(n_{1},n_{2}),
\end{eqnarray}%
where $c$ and $b^\pm_\alpha$ are some constant $k\times k$ matrices, $|n\pm\rangle_{1,2}$ means $|n_{1}\pm 1,n_{2}\rangle$ and $|n_{1},n_{2}\pm 1\rangle$ respectively.
Noting (\ref{quatern}) it is not difficult to have as well%
\begin{eqnarray}
\mathcal{O}|n_{1},n_{2}\rangle &=&(\theta_\beta n_\beta)^{-1}|n_{1},n_{2}\rangle +(\theta_\beta n_\beta)^{-2}C|n_{1},n_{2}\rangle \nonumber\\
&&+\,(\theta_\beta
n_\beta)^{-2}(d^+_\alpha\sqrt{n_\alpha}|n+\rangle_\alpha+d^-_\alpha\sqrt{n_\alpha}|n-\rangle_\alpha) \nonumber\\
&&+\,O(n^{-3})|n_{1},n_{2}\rangle+O(n^{-5/2})|n\pm\rangle \nonumber\\
&&+\,O(n^{-2})|\overline{n_{1},n_{2}}\rangle,
\end{eqnarray}%
where $|\overline{n_{1},n_{2}}\rangle$ stands for any basis states
$|m_{1},m_{2}\rangle$ other than $|n_{1},n_{2}\rangle$ and
$|n\pm\rangle$, $C$ and $d^\pm_\alpha$ are some constant $2k\times
2k$ matrices. So we obtain%
\begin{equation}
\langle n_{1},n_{2}|\mathcal{O}|n_{1},n_{2}\rangle =(\theta_\alpha
n_\alpha)^{-1}+(\theta_\alpha n_\alpha)^{-2}C+O(n^{-3}). \label{O}
\end{equation}

Substituting (\ref{diff}) into (\ref{Stokes}) and using (\ref{O}) we get%
\begin{eqnarray}
\sum_{n_{1},n_{2}=0}^{N_{1},N_{2}}\langle n_{1},n_{2}|\partial _{\alpha }%
\bar{\partial}_{\alpha }\mathcal{O}|n_{1},n_{2}\rangle &=&\frac{N_{1}+1}{%
\theta _{1}}\sum_{n_{2}=0}^{N_{2}}(\langle N_{1}+1,n_{2}|\mathcal{O}%
|N_{1}+1,n_{2}\rangle  \nonumber \\
&&-\langle N_{1},n_{2}|\mathcal{O}|N_{1},n_{2}\rangle )  \nonumber \\
&&+\frac{N_{2}+1}{\theta _{2}}\sum_{n_{1}=0}^{N_{1}}(\langle n_{1},N_{2}+1|%
\mathcal{O}|n_{1},N_{2}+1\rangle  \nonumber \\
&&-\langle n_{1},N_{2}|\mathcal{O}|n_{1},N_{2}\rangle )  \nonumber \\
&=&\frac{N_{1}+1}{\theta _{1}}\sum_{n_{2}=0}^{N_{2}}\{[\theta
_{1}(N_{1}+1)+\theta _{2}n_{2}]^{-1}  \nonumber \\
&&-(\theta _{1}N_{1}+\theta _{2}n_{2})^{-1}+O(N_{1}^{-3})\}  \nonumber \\
&&+\frac{N_{2}+1}{\theta _{2}}\sum_{n_{1}=0}^{N_{1}}\{[\theta
_{1}n_{1}+\theta _{2}(N_{2}+1)]^{-1}  \nonumber \\
&&-(\theta _{1}n_{1}+\theta _{2}N_{2})^{-1}+O(N_{2}^{-3})\}.
\end{eqnarray}%
Then using a simple algebraic trick we have%
\begin{eqnarray}
\textrm{l.h.s.} &=&\frac{N_{1}+1}{-\theta
_{1}}\sum_{n_{2}=0}^{N_{2}}\{\theta _{1}(\theta _{1}N_{1}+\theta
_{2}n_{2})^{-1}[\theta _{1}(N_{1}+1)+\theta
_{2}n_{2}]^{-1}+O(N_{1}^{-3})\}  \nonumber \\
&&+\,(1\leftrightarrow 2)  \nonumber \\
&=&-(N_{1}+1)\sum_{n_{2}=0}^{N_{2}}\{(\theta _{1}N_{1}+\theta
_{2}n_{2})^{-1}[\theta _{1}N_{1}+\theta
_{2}(n_{2}+1)]^{-1}+O(N_{1}^{-3})\}
\nonumber \\
&&+\,(1\leftrightarrow 2)  \nonumber \\
&=&\frac{N_{1}+1}{\theta _{2}}\sum_{n_{2}=0}^{N_{2}}\{[\theta
_{1}N_{1}+\theta _{2}(n_{2}+1)]^{-1}-(\theta _{1}N_{1}+\theta
_{2}n_{2})^{-1}+O(N_{1}^{-3})\}  \nonumber \\
&&+\,(1\leftrightarrow 2)  \nonumber \\
&=&\frac{N_{1}+1}{\theta _{2}}\{[\theta _{1}N_{1}+\theta
_{2}(N_{2}+1)]^{-1}-(\theta
_{1}N_{1})^{-1}+(N_{2}+1)O(N_{1}^{-3})\}
\nonumber \\
&&+\,(1\leftrightarrow 2).
\end{eqnarray}%
Now let $N_{1}=N_{2}=N$, we will obtain%
\begin{eqnarray}
\textrm{l.h.s.} &=&\frac{N+1}{-\theta _{2}}\left[ \frac{\theta _{2}(N+1)}{%
\theta _{1}(\theta _{1}+\theta _{2})N^{2}}+O(N^{-2})\right]
+(1\leftrightarrow 2)  \nonumber \\
&=&-\frac{1}{\theta _{1}(\theta _{1}+\theta _{2})}-\frac{1}{\theta
_{2}(\theta _{1}+\theta _{2})}+O(N^{-1})  \nonumber \\
&=&-\frac{1}{\theta _{1}\theta _{2}}+O(N^{-1}).
\end{eqnarray}%
So the topological charge%
\begin{eqnarray}
Q &=&\frac{\theta _{1}\theta _{2}}{2}\mathrm{Tr}_\mathcal{F}\partial _{\alpha }\bar{%
\partial}_{\alpha }\mathrm{Tr}\mathcal{O}  \nonumber \\
&=&\frac{\theta _{1}\theta _{2}}{2}\mathrm{Tr}\sum_{n_{1},n_{2}=0}^{\infty
}\langle n_{1},n_{2}|\partial _{\alpha }\bar{\partial}_{\alpha }\mathcal{O}%
|n_{1},n_{2}\rangle  \nonumber \\
&=&\theta _{1}\theta _{2}k\lim_{N\rightarrow \infty
}\sum_{n_{1},n_{2}=0}^{N}\langle n_{1},n_{2}|\partial _{\alpha }\bar{\partial%
}_{\alpha }\mathcal{O}|n_{1},n_{2}\rangle  \nonumber \\
&=&-k.
\end{eqnarray}

Then the $\theta _{2}<0$ case is of no difficulty. The only thing
we have to do is to replace $\theta _{2}$ with $-\theta _{2}$ in
the above discussion for $\theta _{2}>0$, except that
(\ref{diff2}) will become
\begin{eqnarray} \langle n_{1},n_{2}|\partial_{2}\mathcal{O}_{2}|n_{1},n_{2}\rangle
&=&\langle n_{1},n_{2}|\frac{\bar{z}_{2}\mathcal{O}_{2}-\mathcal{O}_{2}\bar{z}_{2}}{\theta_{2}}|n_{1},n_{2}\rangle \nonumber\\
&=&(-\theta_2)^{-1/2}(\sqrt{n_2+1}\langle
n_{1},n_{2}|\mathcal{O}_2|n_{1},n_{2}+1\rangle\\
&&-\sqrt{n_2}\langle
n_{1},n_{2}-1|\mathcal{O}_2|n_{1},n_{2}\rangle). \nonumber
\end{eqnarray}
But the final result is not dependent on $\theta _{2}$, so we also
obtain $Q=-k$.

\section*{Acknowledgments}

We would like to thank Jian Dai for useful discussions.

\appendix

\section{Non-singularity of the Operator $\mathcal{O}$ in (\ref{O by f})}
\label{append}
\newtheorem{theorem}{Theorem}
\newtheorem{propo}{Proposition}

Suppose a hermitian operator $O$, acting on a vector $|v\rangle$
in the Hilbert space $\mathcal{H}$, is bounded below but not
bounded above. First we present:
\begin{theorem}\label{min}
The admissible vector $|v\rangle$ which minimize the expression
$\langle v|O|v\rangle$ under the normalization condition $\langle
v|v\rangle=1$ is an eigenvector $|v_0\rangle$ of $O$; the minimum
value $\lambda_0=\langle v_0|O|v_0\rangle$ is the minimum
eigenvalue of $O$. If we impose not only the normalization
condition, but also the orthogonality condition
\begin{equation}
\langle v|v_0\rangle=0,
\end{equation}
then the solution is again an eigenvector $|v_1\rangle$ of $O$;
the minimum value $\lambda_1=\langle
v_1|O|v_1\rangle\geq\lambda_0$ is the second minimum eigenvalue of
$O$. The succussive minimum problems of $\langle v|O|v\rangle$
subject to the normalization condition and to the orthogonality
conditions
\begin{equation}
\langle v|v_i\rangle=0,\quad i=0,1,\cdots,n-1
\end{equation}
define the eigenvectors $v_n$ with $\lambda_n$, the $(n+1)$th
minimum eigenvalues of $O$.
\end{theorem}
It is easy to prove this theorem using the variation of $\langle
v|O|v\rangle/\langle v|v\rangle$. The precondition of this theorem
is that the expectation value $\langle v|O|v\rangle$ does have a
minimum. However, for operators like $f^{-1}=\Delta^\dag\Delta$
which correspond to some differential operators, there is an
existence theorem that guarantees this precondition. We suggest
one to refer to \cite{Hilbert}.

Theorem \ref{min} provides a procedure to obtain the lowest
eigenvalues and the corresponding eigenvectors of $O$. But can
this recurrent procedure give all the eigenvalues and eigenvectors
of $O$? And can these eigenvectors form a complete basis of the
whole Hilbert space? The following theorem gives the answers:
\begin{theorem}\label{inf}
If the eigenvalues $\lambda_n$ obtained by Theorem \ref{min}
become infinite for $n\rightarrow\infty$, then Theorem \ref{min}
gives all the eigenvalues and eigenvectors of $O$. Furthermore,
the eigenvectors $|v_n\rangle$ are complete in $\mathcal{H}$.
\end{theorem}
We will not provide the proof of this theorem here, neither. One
can again refer to \cite{Hilbert} for the proof.

To prove the infinite growth of the eigenvalues is another
problem. But here we need not do this for general operators. We
can directly check that the operator $f^{-1}$ has the property of
infinite growth required by Theorem \ref{inf}. This fact depends
on another theorem in \cite{Hilbert}:
\begin{theorem}
Given $n$ vectors $|u_0\rangle,\cdots,|u_{n-1}\rangle$ and a
normalized vector $|v\rangle$, let $d[u]$ be the minimum value of
$\langle v|O|v\rangle$ under the orthogonality conditions
\begin{equation}\label{ortho}
\langle v|u_i\rangle=0,\quad i=0,1,\cdots,n-1.
\end{equation}
Then $\lambda_n$ is equal to the maximum value of $d[u]$ if the
vectors $|u_0\rangle,\cdots,|u_{n-1}\rangle$ range over all sets
of normalized vectors in $\mathcal{H}$:
\begin{equation}
d[u]\leq\lambda_n,
\end{equation}
where the equality is attained when $|u_0\rangle$, $\cdots$,
$|u_{n-1}\rangle$ are basis vectors of the subspace spanned by
$|v_0\rangle$, $\cdots$, $|v_{n-1}\rangle$.
\end{theorem}
Then we can calculate the lower bound of the eigenvalue
$\lambda_n$ of $f^{-1}$ using this theorem. In this case
$\mathcal{H}=\mathcal{F}\otimes\mathbf{C}^k$, where $\mathcal{F}$
is the Fock space. Choose $|u_0\rangle,\cdots,|u_{n-1}\rangle$ as
all the vectors $|n_1,n_2\rangle\otimes e_i,\ i=1,\cdots,k$ such
that $\theta_\alpha n_\alpha\leq m$. Here $e_i$ are the basis
vectors of $\mathbf{C}^k$:
\begin{equation}
e_i=(\stackrel{1}{0},\cdots,\stackrel{i-1}{0},
\stackrel{i}{1},\stackrel{i+1}{0},\cdots,\stackrel{k}{0})^\dag.
\end{equation}
Obviously, these $|u_0\rangle,\cdots,|u_{n-1}\rangle$ are the
lowest $n$ eigenvectors of $z_\alpha\bar{z}_\alpha\otimes 1_k$. So
we have for any vector $|v\rangle$ which satisfies (\ref{ortho})
\begin{equation}
\langle v|z_\alpha\bar{z}_\alpha|v\rangle\geq m.
\end{equation}
Here we only consider the $\theta_2>0$ case, since in the
$\theta_2<0$ case our discussion needs very few modifications. It
is straightforward to obtain for large $m$
\begin{eqnarray}
\langle v|f^{-1}|v\rangle &=&\langle
v|z_\alpha\bar{z}_\alpha|v\rangle+\langle v|z_\alpha
B_\alpha^\dag|v\rangle+\langle v|B_\alpha\bar{z}_\alpha|v\rangle \nonumber\\
&&+\,\langle v|B_\alpha B_\alpha^\dag|v\rangle+\langle
v|II^\dag|v\rangle \nonumber\\
&\geq &\langle v|z_\alpha\bar{z}_\alpha|v\rangle-2|\langle
v|B_\alpha\bar{z}_\alpha|v\rangle|+\langle v|B_\alpha
B_\alpha^\dag|v\rangle \nonumber\\
&\geq &\langle v|z_\alpha\bar{z}_\alpha|v\rangle-2(\langle
v|B_\alpha B_\alpha^\dag|v\rangle\langle
v|z_\alpha\bar{z}_\alpha|v\rangle)^{1/2}+\langle v|B_\alpha
B_\alpha^\dag|v\rangle \nonumber\\
&=&(\langle v|z_\alpha\bar{z}_\alpha|v\rangle^{1/2}-\langle
v|B_\alpha B_\alpha^\dag|v\rangle^{1/2})^2 \nonumber\\
&\geq &(m^{1/2}-\lambda^{1/2})^2,
\end{eqnarray}
where $\lambda$ is the largest eigenvalue of $B_\alpha
B_\alpha^\dag$, and in the third step we have used the Schwarz
inequality. This equation implies
\begin{equation}
\lambda_n\geq d[u]\geq(m^{1/2}-\lambda^{1/2})^2,
\end{equation}
so the infinite growth of $\lambda_n$ is proved.

So far, by Theorem \ref{inf}, we have obtained:
\begin{propo}
The operator $f^{-1}$ is diagonalizable in the usual sense.
\end{propo}
As we have known, $f^{-1}$ is positive definite \cite{Reviewa}, so
it can be naturally inverted in the diagonal representation. This
leads to:
\begin{propo}
$f$ is a bounded operator on $\mathcal{H}$.
\end{propo}
A bounded operator is non-singular in any representation.

Then, let us consider the projection operator $P=\Delta
f\Delta^\dag$ on $\mathcal{F}\otimes\mathcal{C}^{N+2k}$. By
definition, a projection operator is bounded. The product of two
bounded operators is also a bounded operator. So we have
\begin{propo}
$\mathcal{O}=(2-b^\dag P b)f$ is a bounded operator on
$\mathcal{F}\otimes\mathcal{C}^{2k}$.
\end{propo}
That is enough.

\end{document}